\documentclass[twocolumn]{aastex631}

\usepackage{comment}

\usepackage{graphicx}	
\usepackage{amsmath}	
\usepackage{amssymb}	
\usepackage[para,flushleft]{threeparttable} 
\usepackage{epstopdf}
\usepackage{textgreek}
\usepackage{comment}
\usepackage{multirow}

\usepackage{xcolor, fontawesome}
\definecolor{twitterblue}{RGB}{64,153,255}

\hypersetup{colorlinks,linkcolor={red!60!black},citecolor={blue!60!black},urlcolor={blue!60!black}}

\usepackage{natbib,twoopt}
\bibpunct{(}{)}{;}{a}{}{,} 
\makeatletter
\newcommandtwoopt{\citeads}[3][][]{\href{http://ui.adsabs.harvard.edu/\#abs/#3}%
{\def\hyper@linkstart##1##2{}%
\let\hyper@linkend\@empty\citealp[#1][#2]{#3}}}
\newcommandtwoopt{\citepads}[3][][]{\href{http://ui.adsabs.harvard.edu/\#abs/#3}%
{\def\hyper@linkstart##1##2{}%
\let\hyper@linkend\@empty\citep[#1][#2]{#3}}}
\newcommandtwoopt{\citetads}[3][][]{\href{http://ui.adsabs.harvard.edu/\#abs/#3}%
{\def\hyper@linkstart##1##2{}%
\let\hyper@linkend\@empty\citet[#1][#2]{#3}}}
\newcommandtwoopt{\citeyearads}[3][][]%
{\href{http://ui.adsabs.harvard.edu/\#abs/#3}
{\def\hyper@linkstart##1##2{}%
\let\hyper@linkend\@empty\citeyear[#1][#2]{#3}}}
\makeatother

\graphicspath{{./}{figures/}}

\received{\today}

\shorttitle{Revisiting the space weather environment of Proxima Centauri b} 
\shortauthors{Garraffo et al.}

\begin{document}

\title{Revisiting the space weather environment of Proxima Centauri b}


\author[0000-0002-8791-6286]{Cecilia Garraffo}
\affiliation{Center for Astrophysics \text{\textbar} Harvard \& Smithsonian, 60 Garden Street, Cambridge, MA 02138, USA}

\author[0000-0001-5052-3473]{Juli\'{a}n D. Alvarado-G\'{o}mez}
\affiliation{Leibniz Institute for Astrophysics Potsdam, An der Sternwarte 16, 14482 Potsdam, Germany}

\author[0000-0003-3721-0215]{Ofer Cohen}
\affiliation{Lowell Center for Space Science and Technology, University of Massachusetts Lowell, 600 Suffolk Street, Lowell, MA 01854, USA}

\author[0000-0002-0210-2276]{Jeremy J. Drake}
\affiliation{Center for Astrophysics \text{\textbar} Harvard \& Smithsonian, 60 Garden Street, Cambridge, MA 02138, USA}

\begin{abstract}

Close-in planets orbiting around low-mass stars are exposed to intense energetic photon and particle radiation and harsh space weather.
We have modeled such conditions for Proxima Centauri~b \citep{Garraffo.etal:16b}, a rocky planet orbiting in the habitable-zone of our closest neighboring star, finding a stellar wind pressure three orders of magnitude higher than the solar wind pressure on Earth. At that time, no Zeeman-Doppler observations of the surface magnetic field distribution of Proxima Cen were available and a proxy from a star with similar Rossby number to Proxima was used to drive the MHD model.
Recently, the first ZDI observation of Proxima Cen became available \citepads{2021MNRAS.500.1844K}.  
We have modeled Proxima b's space weather using this map and compared it with the results from the proxy magnetogram. We also computed models for a high-resolution synthetic magnetogram for Proxima b generated by a state-of-the-art dynamo model.
The resulting space weather conditions for these three scenarios are similar with only small differences found between the models based on the ZDI observed magnetogram and the proxy.
We conclude that our proxy magnetogram prescription based on Rossby number is valid, and provides a simple way to estimate stellar magnetic flux distributions when no direct observations are available. Comparison with models based on the synthetic magnetogram show that the exact magnetogram details are not important for predicting global space weather conditions of planets, reinforcing earlier conclusions that the large-scale (low-order) field dominates, and that the small-scale field does not have much influence on the ambient stellar wind. 

\end{abstract}



\section{Introduction} 
\label{sec:intro}

Extensive research has been directed toward understanding the conditions of close-in planets orbiting M dwarfs. These planets are by far the most abundant kind of detected exoplanet orbiting in the temperature-based definition of the habitable zone (HZ). Due to the low luminosity of M dwarfs, their HZ resides very close to the host star (e.g.~Kopparapu~et al.~\citeyearads{2013ApJ...770...82K}, \citeyearads{2014ApJ...787L..29K} \citeads{2016PhR...663....1S}). Low mass stars are typically magnetically more active than higher mass stars, and remain active for much longer (e.g.~\citeads{Reiners.Basri:08},
\citeads{2011ApJ...743...48W}, \citeads{Jackson.etal:12}, \citeads{Cohen.Drake:14}, \citeads{2019ApJ...871..241D}). The associated coronal and chromospheric integrated high-energy radiation can evaporate planetary atmospheres and poses a risk for close-in exoplanets.  In addition, the pressure of the stellar wind also scales with magnetic activity (e.g., as $\dot{M}_{\bigstar} \propto L^{1.34}_{\rm X}$, \citeads{2005ApJ...628L.143W}, \citeads{Vidotto.etal:14} ,\citeads{Garraffo.etal:15a}, \citeads{Garraffo.etal:17}) and therefore close-in planets are expected to experience stronger stellar winds for longer evolutionary timescales, putting their atmospheres at risk of being stripped.

Proxima Centauri b (Proxima b hereafter) is a rocky planet orbiting in the ``habitable zone" of Proxima Centauri, our closest neighboring star at only 1.3 parsecs from Earth \citepads{2016Natur.536..437A}.
Detailed and realistic magnetohydrodynamic (MHD) simulations predicted that Proxima~b should experience stellar wind pressures four orders of magnitude larger than the solar wind pressure experienced at Earth, together with strong variations of this pressure on timescales as short as a day \citep{Garraffo.etal:16b}. Such simulations also predicted that planets around M dwarfs like Proxima~b will suffer from intense Joule heating \citep{2014ApJ...790...57C}, severe atmospheric loss (\citeads{2017ApJ...837L..26D}, \citeads{2017ApJ...844L..13G}), and transitions between sub- and super-Alfv\'enic wind conditions on timescales as short as a day  \citep{2014ApJ...790...57C, Garraffo.etal:17}. 

The MHD wind simulations on which this work is based were driven by the  magnetic fields on the simulated stellar surface (the inner boundary condition). While estimates of the total magnetic flux exist for a number of stars, most stars are either too faint or have surface projected rotation velocities, $v \sin i$, too small, or both, to allow observations of the distribution of the field through the Zeeman-Doppler Imaging (ZDI) technique. 

To get around the lack of ZDI data for particular stars of interest, theoretical arguments have been made to extrapolate surface magnetic structure observed on one star to another when no observations are available. This criterion uses similarities in either spectral type or in the Rossby number---the ratio of rotation period to convective turnover time commonly used as a simple dynamo number \citep[e.g.,][]{Noyes.etal:84,Wright.etal:11}---as an indicator for the large scale structure of the magnetic field.
This approach was adopted by \citet{Garraffo.etal:16b}, who used a magnetogram for the M5~V star GJ~51 from \citet{2010MNRAS.407.2269M} as a proxy for the surface field of Proxima.

Recenty, \citetads{2021MNRAS.500.1844K} reported the first ZDI reconstruction of the large-scale magnetic field of Proxima Cen and provided a rough estimate of its stellar wind stucture based on a potential field approximation. The ZDI observations revealed a relatively simple field geometry, with a dominant $135$~G dipole component displaced by $\sim$\,51$^{\circ}$ with respect to the stellar rotation axis. Due to the extremely slow rotation of Proxima ($P_{\rm rot} \simeq 89$~d, $v\sin(i) \simeq 0.06$~km~s$^{-1}$) and the sparse phase coverage of the observations ($10$~spectropolarimetric exposures covering one rotation), the resulting ZDI map contains very limited information on other components of the surface field. In contrast, state-of-the-art dynamo models tailored to fully-convective stars such as Proxima Cen predict much more complex field distributions, with some solutions producing global-scale mean fields restricted to a single hemisphere (\citeads{2016ApJ...833L..28Y}, \citeads{2020ApJ...902L...3B}). Still, as illustrated by \citetads{2015ApJ...813L..31Y}, due to cancellation effects and limited spatial resolution, it is expected that a complex magnetic field configuration appears much simpler when reconstructed using ZDI.  
While the small-scale magnetic field geometry can affect stellar X-ray emission, stellar winds were shown to be dominated by the large-scale structure of the magnetic field: dipole, quadrupole, and octupole modes, in order of importance \citep{Garraffo.etal:13, Garraffo.etal:18,See.etal:19, See.etal:20}. It has been on this basis that many space weather simulations were run with relatively low resolution ZDI maps \citep{2014ApJ...790...57C, Garraffo.etal:17, AlvaradoGomez.etal:16, AlvaradoGomez.Garraffo.etal:19}. 

The present availability of an observed magnetogram \citepads{2021MNRAS.500.1844K}, a proxy magnetogram \citep{Garraffo.etal:16b}, and a  synthetic magnetogram \citep{2016ApJ...833L..28Y} for Proxima Centauri provides a unique opportunity to compare the geometry of the three, as well as wind model solutions and space environment conditions derived from them. Comparing the winds obtained from these three magnetograms is also important to validate assumptions  made and proxy magnetograms adopted when no direct observations are available to drive simulations for a given system. 

In this work, we re-assess the space weather on Proxima~b in the light of the three new magnetograms. Section.~\ref{sec:proxima} contains an overview of the Proxima Cen system. We compare the magnetogram structures in Sec.~\ref{s:bfield}, and in Sec.~\ref{sec:model}, we describe our numerical methods. In Sec~\ref{sec:results}, we present the resulting space weather conditions on the planet. A summary and conclusions of our work are presented in Sec.~\ref{s:summary}.

\begin{figure}
\center 
\includegraphics[trim = 0.7cm 0.4cm 0.6cm 0.4cm, clip=true,width=.5\textwidth]{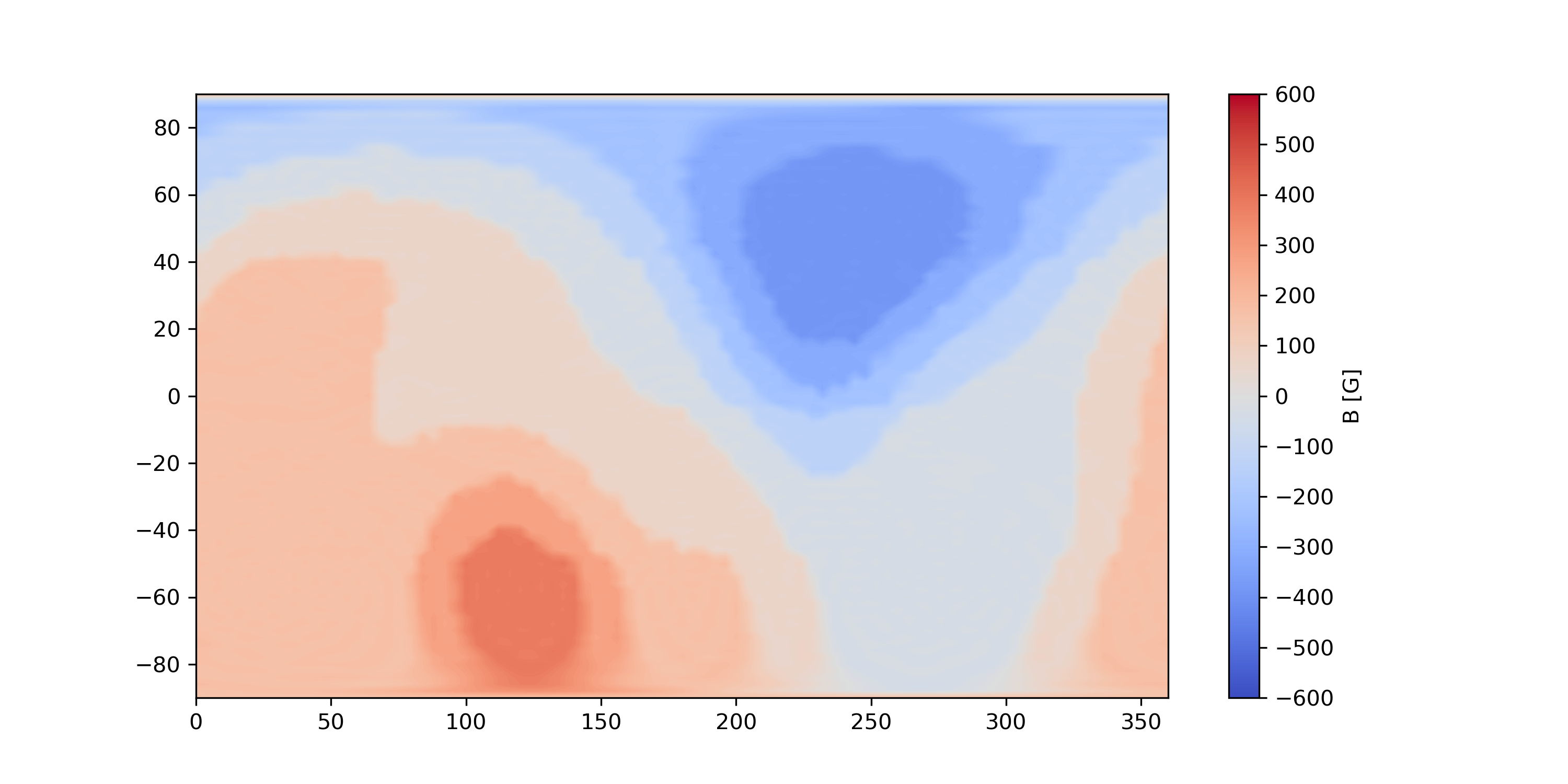}
\includegraphics[trim = 0.7cm 0.4cm 0.6cm 0.4cm, clip=true,width=0.5\textwidth]{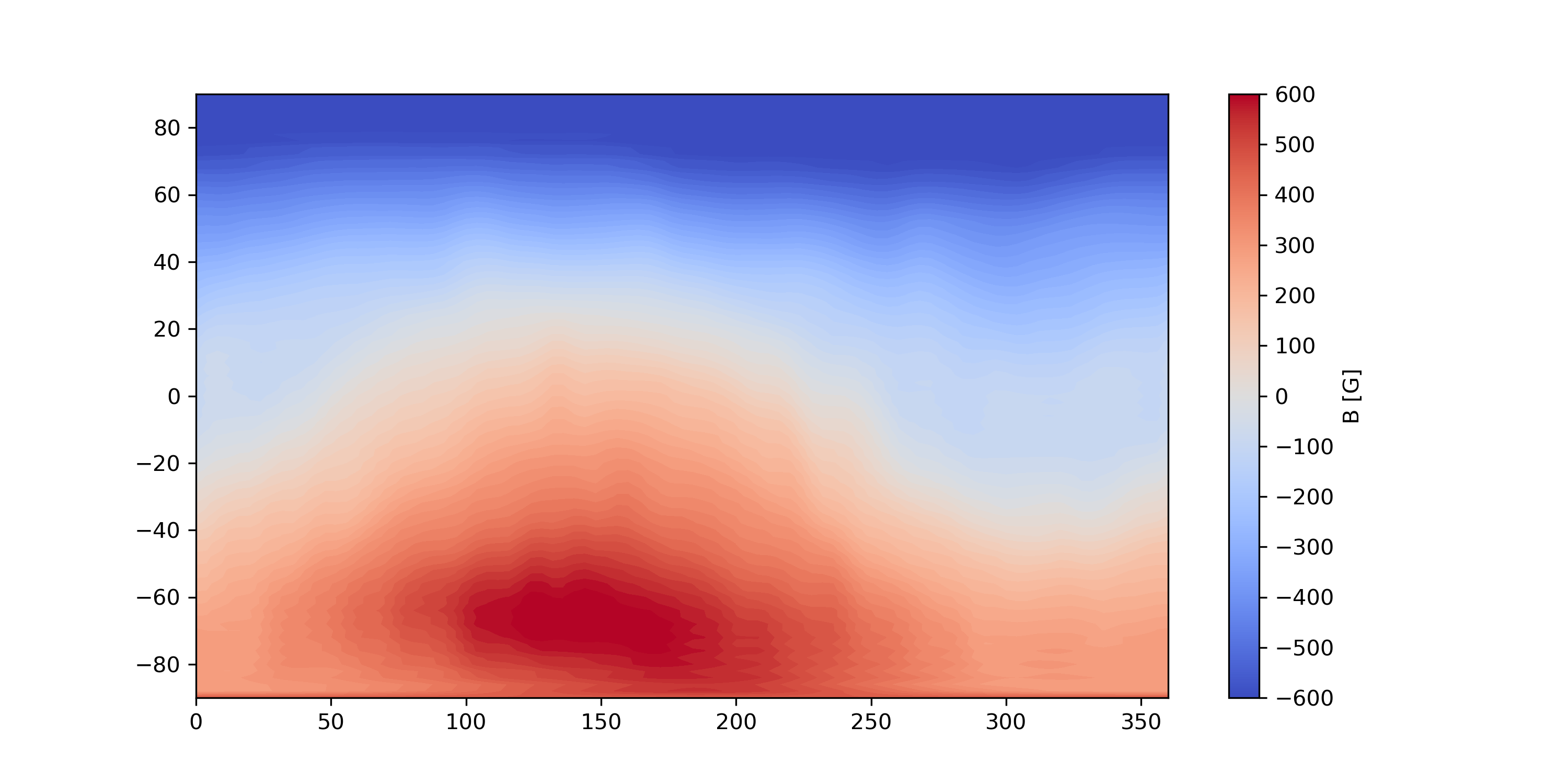}
\includegraphics[trim = 0.7cm 0.4cm 0.6cm 0.4cm, clip=true,width=0.5\textwidth]{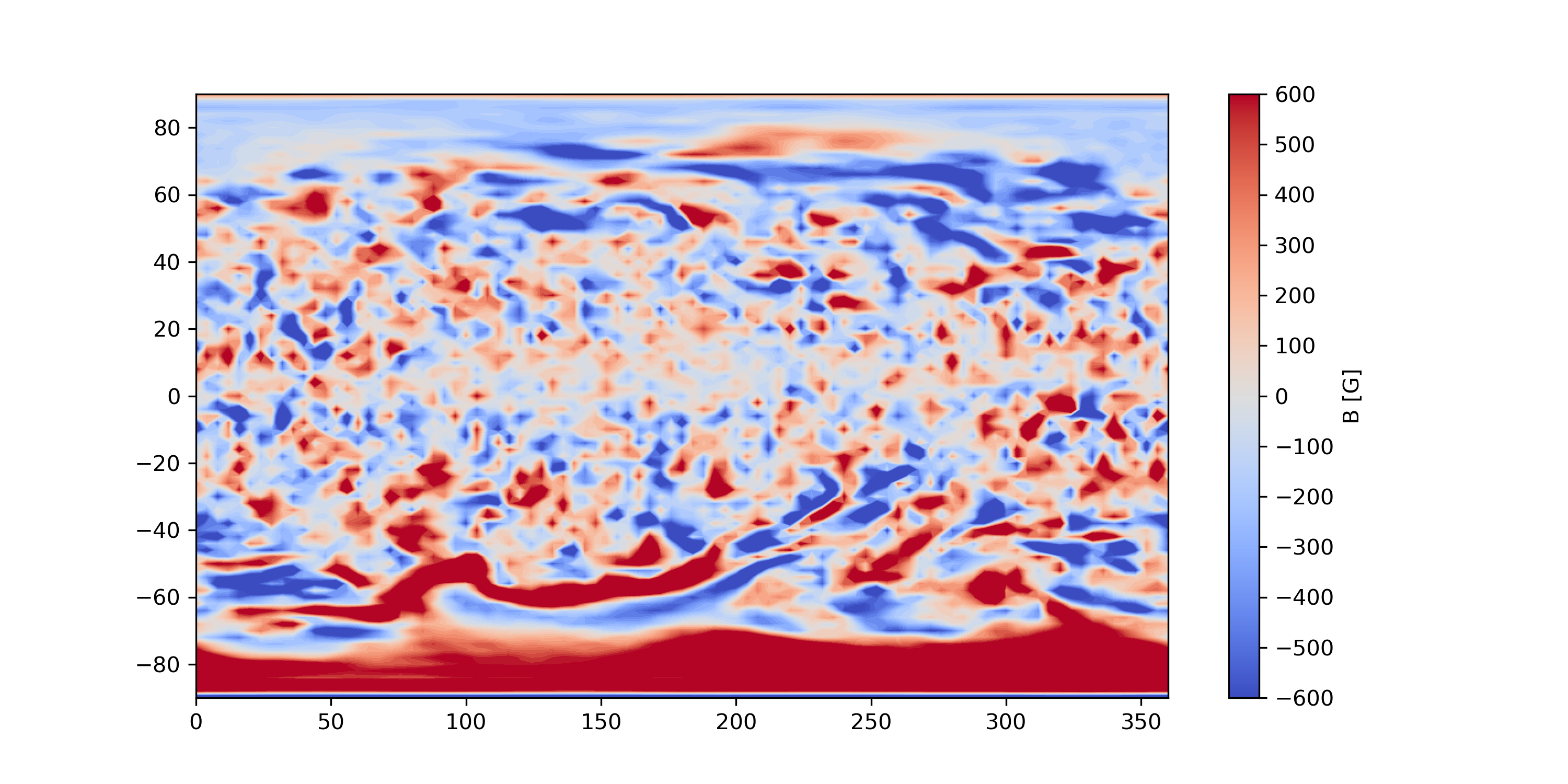}
\caption{The different Proxima magnetograms used in this study. From top to bottom: The observed ZDI magnetogram from \citeads{2021MNRAS.500.1844K}, the proxy magnetogram for Proxima Cen from \citeads{Garraffo.etal:16b}, and the synthetic magnetogram from the dynamo simulations of \citeads{2016ApJ...833L..28Y}.}
\label{fig:magnetograms}
\end{figure}

\section{The Proxima Centauri System} 
\label{sec:proxima}

Proxima Centauri is a late M dwarf (M5.5) with a mass of 0.122 $M_{\odot}$, a radius of 0.154 $R_{\odot}$ \citepads{2016Natur.536..437A}, and a rotation period of 83 days \citep{Kiraga.Stepien:07}. Its age has been estimated to be 4.85~Gyr \citep{2003A&A...397L...5S, 2007AcA....57..149K, 2016Natur.536..437A}. Proxima Centauri~b, a rocky exoplanet orbiting in its habitable zone, has an orbital radius of just 0.049~au, 20 times smaller than Earth's orbit. Its mass has been estimated to be at least 1.7 $M_{\oplus}$ (\citeads{2020A&A...639A..77S})  and its equilibrium temperature to be 234~K \citep{2016Natur.536..437A}, comparable to the Earth's one of 255~K. 

Proxima~b has become an icon for potential habitable planets. Its irradiation history has been closely studied in order to assess the climate evolution of the planet (e.g.~\citeads{2016A&A...596A.111R}, \citeads{2016A&A...596A.112T}).  Its quiescent space weather conditions have also been modeled in detail and show harsh conditions are expected at the location of the planet, potentially leading to atmospheric stripping, heating and evaporation \citep{Garraffo.etal:16b}.  

A magnetic cycle of approximately  seven years was reported by \citepads{2017MNRAS.464.3281W}. However, this variability has negligible effects on space weather conditions that have been shown to be very corrosive for a wide range of magnetic activity levels, reaching wind pressures 4 orders of magnitude higher than the solar wind pressure experienced at Earth \citep{Garraffo.etal:16b, Garraffo.etal:17}.  

Proxima is a flare star (e.g.~\citeads{2011A&A...534A.133F}, \citeads{2019ApJ...884..160V}), and its circumstellar conditions are expected to be even harsher when considering transient effects \citep{2019ApJ...884L..13A, 2020ApJ...895...47A}. 
There are also presently two candidate coronal mass ejections from Proxima, one based on X-ray absorption seen by the {\it Einstein} satellite (\citeads{Haisch.etal:83},  \citeads{2019ApJ...877..105M}) and one based on detection of a Type IV radio burst coincident with a white light flare \citepads{Zic.etal:20}.

The presence of a second planet, Proxima~c, has been examined \citepads{2020SciA....6.7467D}. If confirmed, this planet would have an  estimated mass of $\sim6-7~M_\Earth$  and a circular orbit of approximately $1.44$~au (\citeads{2020A&A...635L..14K}, \citeads{2020RNAAS...4...46B}, \citeads{2020A&A...638A.120G}, \citeads{2020RNAAS...4...86B}). An additional short-period sub-Earth has recently been detected \citep{Faria.etal:22}, with $M\sin i\simeq0.26~M_\Earth$, which becomes the innermost planet in the system at an orbital distance of $\sim0.029$~au.

\section{The magnetic field distribution of Proxima}
\label{s:bfield}

Magnetic fields on the surfaces of late-type stars are thought to be the key ingredient driving their winds. Since the energy to drive the wind comes from the magnetic field, it is assumed that the greater the magnetic flux is, the stronger the winds will be. However, it is now recognized that the distribution of the magnetic field over the stellar surface is also a significant factor that plays into the structure of the resulting stellar wind \citep{Vidotto.etal:14, Garraffo.etal:15, Reville.etal:15}. 

Zeeman-splitting observations can be used to estimate the stellar magnetic flux. \cite{Reiners.Basri:08} have estimated the average field of Proxima Centauri to be $\sim 600G$. Observations of the magnetic field geometry on the stellar surface are instead sparse since they require ZDI, which is a more demanding method that requires the source star to be  comparatively bright and to rotate fast enough for a Doppler shift-modulated magnetic signature to be detected \citep[e.g.,][]{Donati.Landstreet:09}. Realistic MHD simulations of stellar winds are driven by the magnetic map of the stellar surface field (a ``magnetogram''). Therefore, the main limitation on obtaining these maps for stars translates to a limitation in modelling their winds, mass loss, and angular momentum loss, which are all relevant for a number of astrophysical phenomena, like stellar spin down and evolution, and the space weather of exoplanets. 

Stellar magnetic activity is driven by the dynamo processes, fueled by rotation such that faster rotation results in stronger magnetic activity \citep[e.g.][]{Kraft:67,Skumanich:72,Vaiana.etal:81}. Additionally, X-ray observations, in concert with magnetic activity indicators at other wavelengths, reveal a strong correlation between magnetic activity and Rossby number ($Ro = P_{rot}/\tau$, where $P_{rot}$ is rotation period and $\tau$ the convective turnover time; see e.~g., \citeads{Noyes.etal:84}, \citeads{Pizzolato.etal:03},  ~\citeads{2011ApJ...743...48W}).   
It has recently been realised that the distribution of magnetic fields on the stellar surface also seems to be governed by $Ro$ \citep[][see also \citealt{2010MNRAS.407.2269M,2013A&A...549L...5G}]{Garraffo.etal:18}. Using that information, one can choose a suitable representative proxy magnetogram obtained for a star with similar stellar properties and $Ro$ as the star of interest when ZDI observations are not available. That technique has been used several times, allowing simulations of the space enviroment of Proxima Centauri \citep{Garraffo.etal:16b}, TRAPPIST-1 \citep{Garraffo.etal:17}, Barnard's Star \citepads{2019ApJ...875L..12A}, and TOI-700 \citepads{2020ApJ...897..101C}.

The proxy GJ~51 magnetogram employed by \citet{Garraffo.etal:16b} is compared with the newly observed Proxima ZDI map by \citetads{2021MNRAS.500.1844K}, and the synthetic magnetogram generated by the \citetads{2016ApJ...833L..28Y} dynamo model in Figure~\ref{fig:magnetograms}.
We see that the observed and the proxy ones are similar, both in structure and in magnetic field strengths. They are both dominantly dipolar and with a maximum field strength of $\sim$ 600~G. The dynamo simulated magnetogram has, as expected, much higher resolution than the observed ones (note that the proxy magnetogram is also an observed ZDI map for a different star).  There is a lot of small structure and concentrated field. The field strength in the dynamo model has been normalized to a Br-max = 600~G. This has been made not only to make it more comparable to the observed ones, but also because the original Br from the simulation is extracted at R = 0.95 Rstar, and therefore with much larger field strengths due to the greater ambient pressure and concentration of field. The total magnetic flux in each of the magnetograms is $\Phi_T =2.0\times 10^{25}$~Mx (observed), $4.25\times 10^{25}$~Mx (proxy), and $4.4\times 10^{25}$~Mx (dynamo simulation).

\begin{figure*}
\center 
\includegraphics[trim = 0.4cm 1.8cm 10cm 1.3cm, clip=true, width=0.32\textwidth]{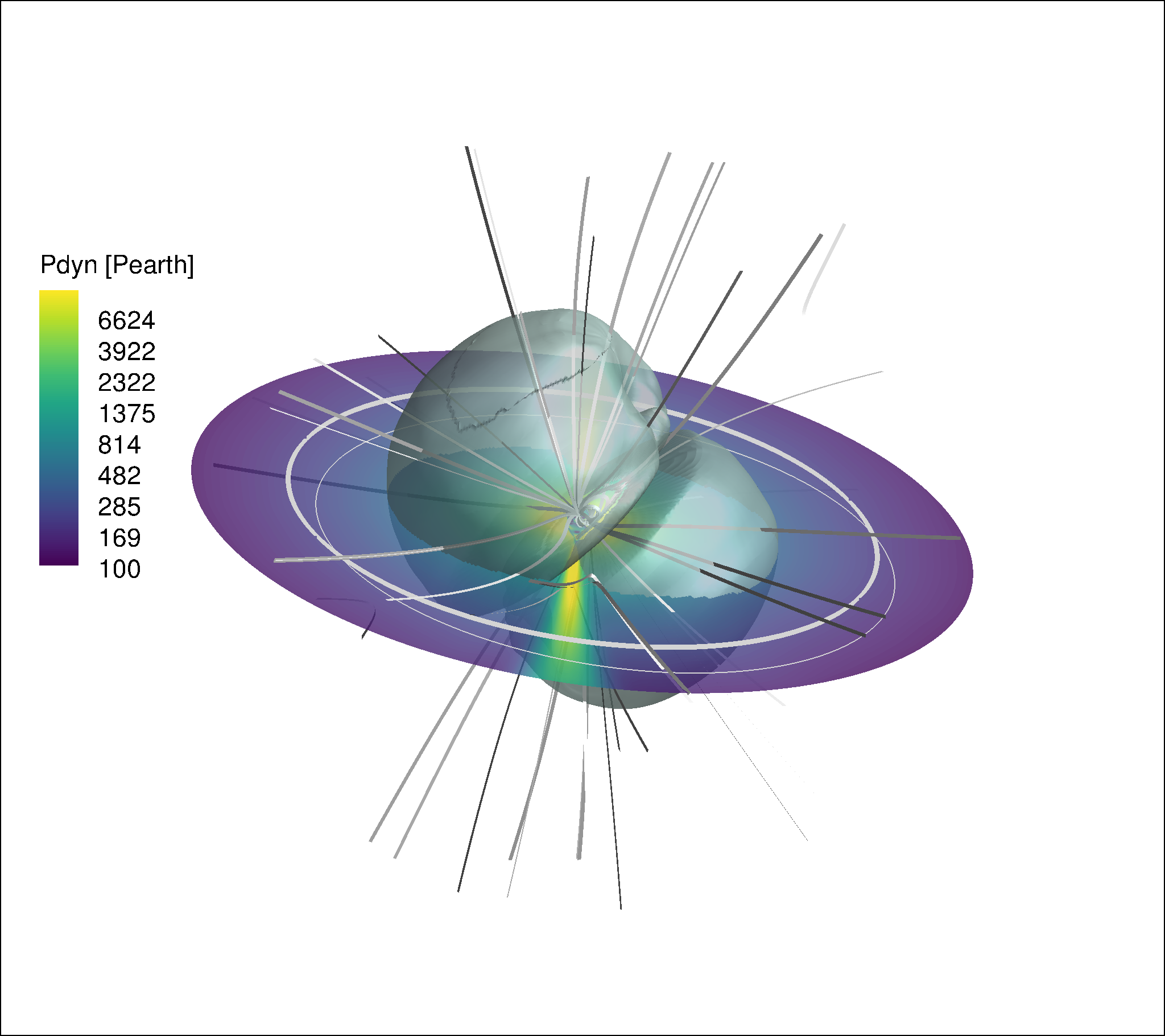}
\includegraphics[trim = 4.4cm 0.4cm 4.5cm 0.5cm, clip=true, width=0.3\textwidth]{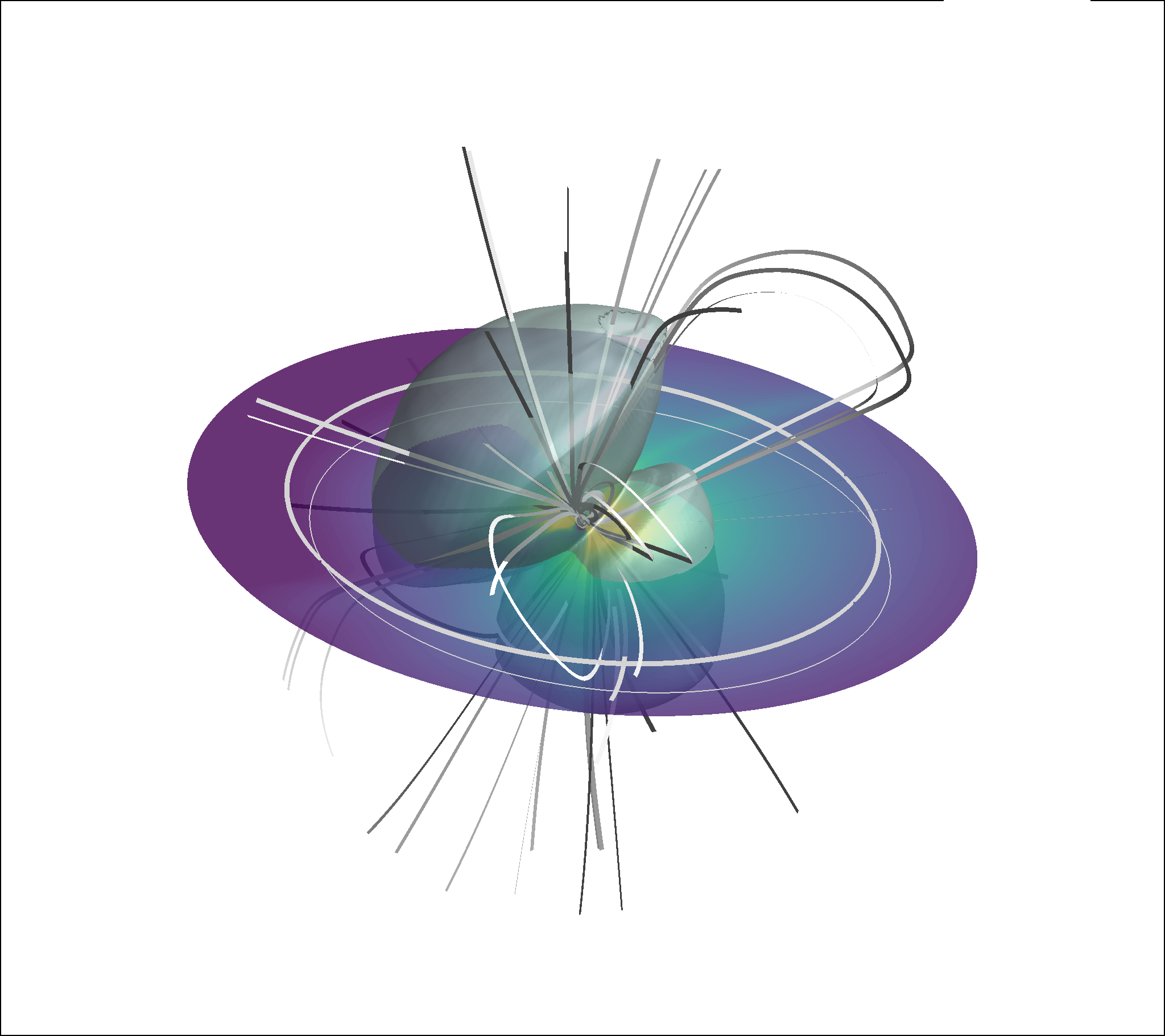}
\includegraphics[trim = 16cm 8cm 0.4cm 1cm, clip=true, width=0.34\textwidth]{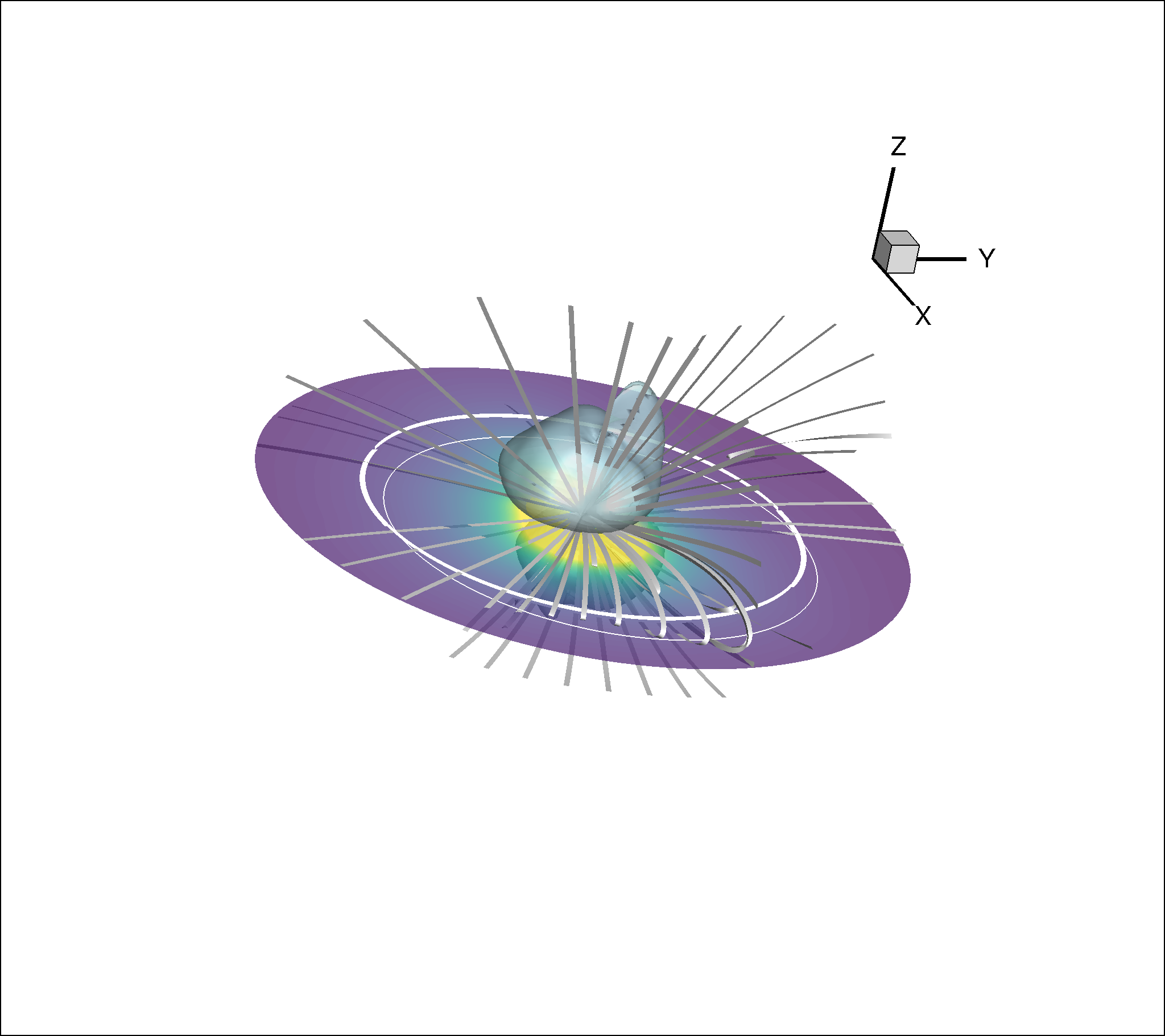}
\caption{Three-dimensional stellar wind simulations of Proxima Centauri driven by the observed ZDI magnetogram (left), the proxy magnetogram (middle), and the synthetic dynamo generated magnetogram (right). The equatorial plane is colored according to dynamic wind pressure normalized to the solar wind pressure at 1 AU. The simulated orbits for Proxima b are in white (the thick line is the circular orbit). The grey shaded surface denotes the Alfv\'en surface. The colored domain has a radius of $110R_\star$ ($16.9 R_\odot$).}
\label{fig:3D}
\end{figure*}

In order to compare the magnetic complexity of the magnetograms, we performed spherical harmonic  decomposition and calculated the average large-scale order, $n_{av}$, as in \cite{Garraffo.etal:16a}, by taking the weighted average of the magnetic multipolar order,
\begin{equation}
    n_{av}=\sum_{n=0}^{n_{max}} \frac{n\Phi_n}{\Phi_T}
\end{equation}
where $n$ is the multipolar order, and $\Phi_n$ is the magnetic flux in each term of the decomposition.
We consider the large-scale structure to be all orders lower than or equal to 7. We find that the dominant large-scale orders are $n_{av}=1.7$, 1.5, and 3.8 for the observed ZDI, the Proxy, and the synthetic magnetograms, respectively. In accord with the appearance of being much more high-order dominated, the model magnetogram is then significantly more complex on average that the observed ones.

\section{Wind Model}
\label{sec:model}

We use the {\it AWSOM} model \citep{2014ApJ...782...81V} to simulate the stellar corona and stellar wind. The model uses the input magnetogram to specify the radial magnetic field distribution and to calculate the potential magnetic field in the whole domain.  This three-dimensional field serves as the initial potential magnetic field in the simulation. The model then solves the non-ideal MHD equations (the conservation of mass, momentum, magnetic induction, and energy), taking into account Alfv\'en wave coronal heating and solar wind acceleration as additional momentum and energy terms. It also accounts for radiative cooling and electron heat conduction. The final steady-state solution is the non-potential, energized corona, and accelerated stellar wind, where the overall structure of the wind solution follows the structure of the input magnetogram field. Numerical validation of the AWSOM/BATSRUS model, including standard numerical tests and grid convergence have been presented in \cite{powell1999, toth2005,toth2012}, and \cite{2014ApJ...782...81V}. Validation of the model against solar and solar wind observations has been presented in \cite[e.g., ][]{2014ApJ...782...81V, Sachdeva2019}. An initial grid refinement is applied with a radially-stretched grid. This creates a grid with a smallest grid size of $\Delta x=0.026R_\star$ near the inner boundary, and a large grid size of $\Delta x=0.5R_\star$ near the outer boundary. All three cases were simulated with an identical grid to remove any possible impact of the grid on the results. We refer the reader to \cite{2014ApJ...782...81V} for a complete description of the model. 

The wind model has two main free parameters that control the solution (other than the input magnetogram). One is the Poynting flux that is provided at the inner boundary. This parameter dictates how much wave energy is supplied at the footpoint of a coronal magnetic field line. The other parameter is related to the proportionality constant that controls the dissipation of the Alfv\'en wave energy into the coronal plasma. For consistency, we use the same set of parameters for all three input magnetograms, where the Poynting flux (per unit magnetic field) value is $1.1\times 10^6~[W~m^{-2}~T^{-1}]$, and the dissipation parameter equals $6\times10^5~[m\sqrt{T}]$. The former is the same value that is typically used for solar simulations, while the latter is slightly  higher due to the fact that the stellar magnetic field of Proxima Centauri is much stronger than that of the Sun \citep{2014ApJ...782...81V,2019ApJ...887...83S}.  

Wind models were computed for each of the three magnetograms and wind conditions were extracted at the putative orbit of Proxima~b. Neither the orbital eccentricity nor the inclination of Proxima~b with respect to the stellar rotation axis, are known. 
We extracted conditions at an orbital inclination of $0\deg$ and for eccentricities of $e=0$ and 0.2. This differs from \citetads{Garraffo.etal:16b}, who examined $10\deg$ and $60\deg$ inclinations, however the difference between conditions at $0\deg$ and $10\deg$ inclinations is negligible.

\begin{figure*}[h!]
\centering
\includegraphics[width=7in]{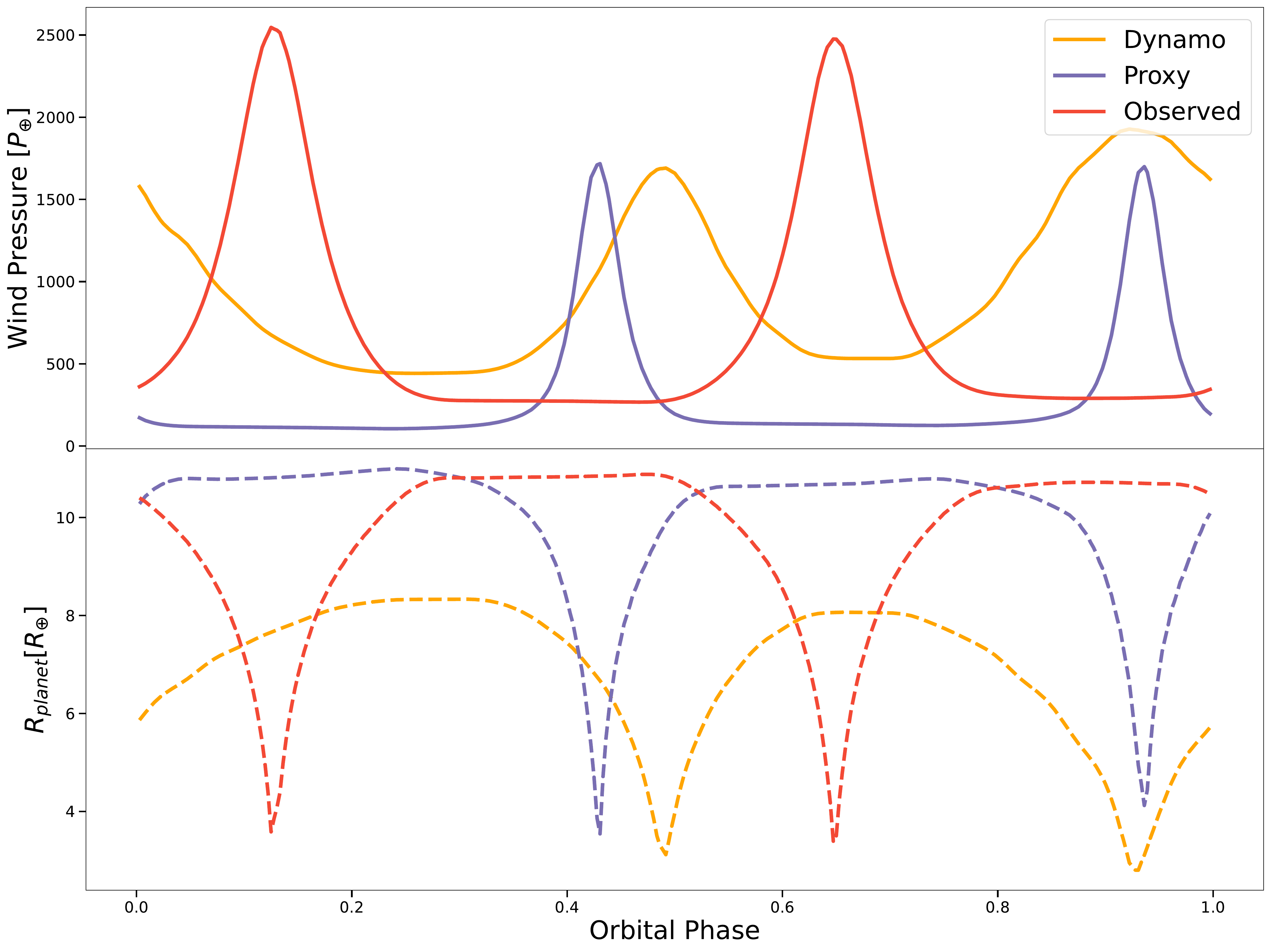}
\caption{Top: dynamic wind pressure normalized to the solar wind pressure at Earth using the proxy map (purple), the Proxima Centauri observed map (red), and the dynamo map (orange). Bottom: magnetospheric standoff distance for the same cases.}
\label{fig:orbits}
\end{figure*}

\section{Results and Discussion} 
\label{sec:results}

Wind model results for the three magnetograms showing the wind dynamic pressure, some selected magnetic field lines, and the Alfv\'en surface are illustrated in Figure~\ref{fig:3D}.

Despite the differences in the magnetograms, the wind solutions in all three scenarios have similar structures although there are some differences in dynamic pressures.  The latter is borne out more clearly in Fig.~\ref{fig:orbits}, showing the dynamic pressure at the orbital distance of Proxima~b as a function of orbital phase for each model wind solution (solid lines). As expected, the results for eccentricities $e=0$ an 0.2 are very similar, and, therefore, are not illustrated. The main differences between the three solutions are due to the different orientations of the dipolar magnetic axis with respect to the orbital plane axis. This can also be seen clearly from Fig.~\ref{fig:3D}. In the case driven by the dynamo-generated magnetogram, the two axes are closely aligned. Therefore, the orbital plane is essentially the same as the current sheet plane. Note that the phases are arbitrarily chosen. In all three cases, the orbit of Proxima~b lies outside the stellar Alfv\'en radius and experiences supersonic wind conditions.

The similar wind solutions from such different magnetograms---observed and proxy versus simulated---does not come as a surprise since, as discussed in Sec.~\ref{sec:intro}, it is the large scale structure (dipole, quadrupole and octupole components) that determines the wind structure. The sizes of the resulting  Alv\'en surfaces are comparable for the three cases: very similar for ZDI and proxy driven simulations, and smaller for the dynamo-simulated scenario. Once again, this latter result is expected when the field is higher-order \citep{Garraffo.etal:16a}. We find that the wind speeds and wind densities and, therefore, the wind pressures are very similar in all cases (see Fig.~\ref{fig:3D} and \ref{fig:orbits}). 

The average wind pressures for the three cases fall in the range 100-300 times the solar wind pressure at at Earth. The variability is similar for the three cases, although more pronounced for the proxy and ZDI magnetic maps. The reason for this is that, in these cases, the orbits experience two crossings of the astrospheric current sheet per orbit (where the density is higher due to the mostly dipolar large-scale magnetic field). This is a consequence of the inclination of the magnetic axis with respect to the rotation axis. For the dynamo driven solution, both axis are aligned and, therefore, the variability is smaller. 

It is important to note that these are steady-state simulations and reflect the quiescent stellar winds for each magnetogram. However, changes in the surface magnetic field of Proxima Centauri are expected to arise from its 7~yr magnetic cycle. Those changes amount to about a factor 2 in the X-ray emission \citep{2017MNRAS.464.3281W}. Based on the empirical relation between unsigned magnetic flux, $\Phi$, and X-ray luminosity, $L_X$, found by \citet{2003ApJ...598.1387P}, $L_X\propto\Phi^
{1.13}$, the expected change in magnetic flux over the cycle is also expected to be approximately a factor of 2. This is similar to the 
difference in total magnetic flux between the observed and proxy magnetograms described in Sect.~\ref{s:bfield}. Magnetic cycle-induced  variations in space weather in the Proxima system are not expected to be severe (see \citeads{2020ApJ...902L...9A}).

In order to get a better idea of how the different wind conditions through the orbit and due to the different magnetograms might affect Proxima~b, we calculate the size of the planetary magnetosphere along its orbit, assuming a planetary surface magnetic field of 0.1~G as estimated by \citet{Zuluaga.Bustamante:18}. We use the conventional equation for the magnetopause standoff distance, $R_{mp}$, that equates the magnetosphere dipolar field magnetic pressure with the wind's dynamic pressure, $P_{wd}$, \citep[e.g., ][]{Kivelson.Russell:95,Gombosi:04},
\begin{equation}
    \frac{R_{mp}}{R_p}=\left[\frac{B_p^2}{4\pi P_{wd}}\right]^{1/6},
\end{equation}
where $R_P$ and $B_p$ are the planetary radius and surface magnetic field strength, respectively.
Fig.~\ref{fig:orbits} (dashed lines) shows the size of the resulting magnetosphere for the three cases. It can be seen that the magnetosphere size changes inversely with the stellar wind pressure, and the overall size of the magnetosphere is not that different for all three cases.

It is of interest to compare the results from the detailed MHD simulations presented here with the potential field-based estimates of conditions by \citetads{2021MNRAS.500.1844K}.
\citetads{2021MNRAS.500.1844K} assumed a constant stellar wind speed, and concluded that stellar wind variations around Proxima Centauri b should be roughly constant. In realistic simulations we find instead that different stellar wind sectors can have quite different conditions, as previously noted in \citetads{Garraffo.etal:16b}. Our MHD wind solution finds quite large wind dynamic pressure variations along the planetary orbit amounting to factors of 3 (the dynamo model magnetogram) to factors of 10 (the observed magnetogram) that would occur on timescales of a day or less.

\citetads{2021MNRAS.500.1844K} suggest that Proxima~b could sustain a magnetosphere of radius 2-$3R_p$, based on planetary parameters from \citeads{2016A&A...596A.111R}. Those authors investigated magnetic fields of $B_p=B_\oplus$ and $0.2B_\oplus$, where $B_\oplus$ corresponds to the field strength of the Earth. The value assumed for the Earth’s magnetic field strength was not stated, but is commonly assumed to be 0.3~G \citep[e.g., ][]{Kivelson.Russell:95,Gombosi:04}. This would correspond to field strengths in the \citetads{2021MNRAS.500.1844K} calculations of 0.3 and 0.06~G, which span the value of 0.1~G assumed here. Here, we find the magnetopause standoff distance that strongly varies with time, with values ranging between approximately $3R_p$ and $11 R_p$, depending on the orbital phase and wind solution.

\section{Summary and Conclusions}
\label{s:summary}

The main conclusion from this work is that the differences between the three magnetogram scenarios explored are at the detail level, and not significant for a description of the global space weather. Such differences are also within the range of expected magnetic-cycle variability of Proxima Centauri. 

The proxy magnetogram criteria used by \cite{Garraffo.etal:16b, Garraffo.etal:17} results in a space weather environment not significantly different than the one resulting from the observed ZDI magnetogram. The reason is that the large-scale magnetic field distribution on the stellar surface is mainly dictated by the stellar rotation period and mass, through the Rossby number prescription.  Based on those stellar parameters one can estimate the magnetic flux and the order of the magnetic field distribution.  This validates the proxy method and justifies the use of a representative ZDI observation for a star for which no ZDI is available. This will allow the community to make advances towards reliably assessing the space weather conditions on a vast number of interesting systems. 

Our results also support the dynamo simulations by \cite{Yadav.etal:15}. The synthetic magnetograms resulting from the dynamo simulations look significantly different to the observed ZDI maps. However, the fact that they lead to a similar global space weather suggests that they capture the important large-scale distribution of the field.

\vspace{0.5in}

We thank the anonymous referee for a very thorough, useful and constructive report.  CG and JJD were supported by NASA contract NAS8-03060 to the {\it Chandra X-ray Center}. Simulation results were obtained using the (open source) Space Weather Modeling Framework, developed by the Center for Space Environment Modeling, at the University of Michigan with funding support from NASA ESS, NASA ESTO-CT, NSF KDI, and DoD MURI. The simulations were performed on NASA's Pleiades cluster (under SMD-17-1330 and SMD-16-6857), as well as on the Massachusetts Green High Performance Computing Center (MGHPCC) cluster.


\bibliographystyle{aasjournal}
\bibliography{Biblio}



\end{document}